%% file: main.tex
\documentclass[sigconf]{acmart}

\AtBeginDocument{%
  \providecommand\BibTeX{{%
    \normalfont B\kern-0.5em{\scshape i\kern-0.25em b}\kern-0.8em\TeX}}}

\copyrightyear{2021}
\acmYear{2021}
\setcopyright{acmcopyright}\acmConference[PREPRINT]{PREPRINT}
\acmBooktitle{ARXIV PREPRINT}

\renewcommand\footnotetextcopyrightpermission[1]{}
\usepackage{amsmath,amsfonts}
\usepackage{algorithmic}
\usepackage{graphicx}
\usepackage{textcomp}
\usepackage{xcolor}
\usepackage{xspace}
\usepackage{pifont}
\usepackage{enumitem}
\usepackage{todonotes}

\newcommand{\mpara}[1]{\smallskip\noindent{\bf #1}}
\newcommand{\spara}[1]{\smallskip\noindent{\bf #1}}

\newcommand{\oursystem}{\textsc{Sofos}\xspace}
\newcommand{\template}{facet\xspace}
\newcommand{\templates}{facets\xspace}

\newcommand{\Q}{Q}
\newcommand{\V}{\mathcal{V}}

\newcommand{\kg}{G}
\newcommand{\T}{F}
\renewcommand{\C}{C}

\newcommand{\lattice}[1]{\mathcal{V}(#1)}
\newcommand{\res}[2]{#1(#2)}

\newcommand{\iris}{\mathcal{I}}
\newcommand{\blanks}{\mathcal{B}}
\newcommand{\literals}{\mathcal{L}}
\newcommand{\vars}{\mathcal{X}}

\newcommand{\varsq}{\vec{X}}

\newcommand{\reals}{\mathbb{R}}

\begin{document}
\fancyhead{}
\title{\oursystem: Demonstrating the Challenges of Materialized View Selection on Knowledge Graphs
}

\author{Georgia Troullinou}
\affiliation{%
  \institution{FORTH-ICS}
   \country{Greece}
 }
\email{troulin@ics.forth.gr}
\author{Haridimos Kondylakis}
\affiliation{%
  \institution{FORTH-ICS}
 \country{Greece}
 }
\email{kondylak@ics.forth.gr}
\author{Matteo Lissandrini}
\affiliation{%
  \institution{Aalborg University}
  \country{Denmark}
 }
\email{matteo@cs.aau.dk}
\author{Davide Mottin}
\affiliation{%
  \institution{Aarhus University}
 \country{Denmark}
 }
\email{davide@cs.au.dk}

\begin{abstract}
Analytical queries over RDF data are becoming prominent as a result of the proliferation of knowledge graphs.
Yet, RDF databases are not optimized to perform such queries efficiently, leading to long processing times.
A well known technique to improve the performance of analytical queries is to exploit \emph{materialized views}.
Although popular in relational databases, view materialization for RDF and SPARQL has not yet transitioned into practice, due to the non-trivial application to the RDF graph model.
Motivated by a lack of understanding of the impact of view materialization alternatives for RDF data, we demonstrate \oursystem, a system that implements and compares several cost models for view materialization.
\oursystem is, to the best of our knowledge, the first attempt to adapt cost models, initially studied in relational data, to the generic RDF setting, and to propose new ones, analyzing their pitfalls and merits.
{\oursystem} takes an RDF dataset and an analytical query for some \template in the data, and compares and evaluates alternative cost models, displaying statistics and insights about time, memory consumption, and query characteristics.

\end{abstract}


\maketitle
\sloppy

\input{sections/1.introduction}

\input{sections/2.related}

\input{sections/4.solution}

\input{sections/5.scenario}

\vspace*{-6pt}
\begin{acks}
Matteo Lissanrini is supported by the EU’s H2020 research and innovation programme under the Marie Skłodowska-Curie grant agreement No 838216. This research project was supported by the Hellenic Foundation
for Research and Innovation (H.F.R.I.) under the “2nd Call for H.F.R.I. Research Projects to support Post-Doctoral Researchers” (iQARuS Project No 1147).
\end{acks}

\clearpage
\bibliographystyle{ACM-Reference-Format}
\bibliography{bibliography}

\end{document}

%% file: sections/1.introduction.tex
\section{Introduction}\label{sec:introduction}

\begin{figure}
    \centering
    \includegraphics[width=.9\columnwidth]{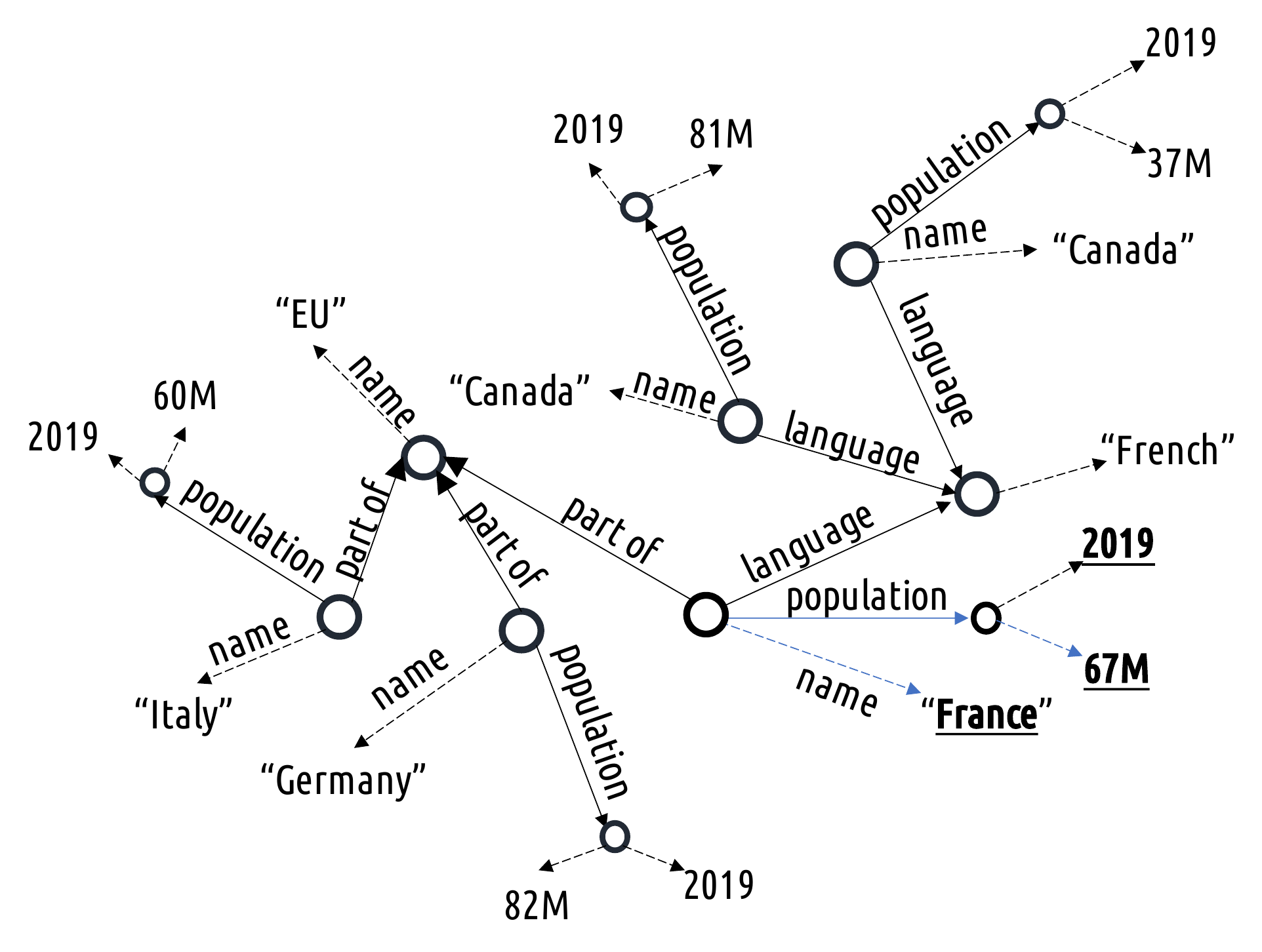}

    \caption{An example Knowledge Graph.}
    \label{fig:graph}

\end{figure}

Companies of all types and sectors, such as Amazon, Google, Bosh, and Zalando, use the graph model to represent and store their enterprise knowledge bases~\cite{noy2019industry,Schmid2019UsingKG}.
Moreover, large knowledge repositories are now available with a wide range of information in many different domains -- DBpedia and WikiData are two notable examples.
Most of this knowledge is available as RDF datasets~\cite{RDF} through SPARQL endpoints~\cite{Bonifati2019}, organized as \emph{knowledge graphs} (KGs).
In KGs like the one in Figure~\ref{fig:graph}, nodes represent entities and edges represent relationships and attributes.
KGs allow storing a wide range of heterogeneous, factual, and statistical information that forms a valuable asset for businesses, organizations, and individuals.

As more data is stored in KGs, there is an increasing need to answer more complex queries~\cite{AnalyticsSPARQL,noy2019industry}.
However, in SPARQL query processing, the research mainly focuses on queries that identify nodes and edges satisfying some specific conditions (e.g., entities by name, friends of friends, or product categories)~\cite{watdiv,lubm,Bonifati2019}.
\begin{example}
Consider a KG like DBpedia or WikiData storing for each country the list of official languages and the number of people speaking that language in that country.
This data can be used to answer analytical queries like ``in  how many countries is French an official language?'' or ``what is the total amount of French-speaking population in the American continent?''.
\end{example}
\noindent Given the growing importance of KGs as knowledge repositories, there is a need for effective \emph{analytical query} answering to extract relevant insights from the data~\cite{colazzo2014rdf,AnalyticsSPARQL,olapcubeRDF}.

The study of analytical queries (i.e., OLAP) over relational systems has attracted substantial attention in the past decades~\cite{niemi2001constructing} and recently, different methodologies have also been proposed in the context of KGs~\cite{colazzo2014rdf,gur2017geosemolap}.
Nonetheless, obtaining answers to analytical queries is usually time-consuming and prohibitively expensive for most RDF data-management systems~\cite{AnalyticsSPARQL}.
A technique to improve the performance of analytical queries is view materialization~\cite{olapcube}.
View materialization precomputes and stores the results of analytical queries offline to serve new incoming queries faster.
Nonetheless, this requires the system to select which views to materialize.
In addition, the intricacies of the RDF model, e.g., complex schema, entailment, and blank nodes, further complicate the direct adoption of techniques proposed for the relational data.

A recent work~\cite{olapcubeRDF} applies an approach designed for relational OLAP~\cite{olapcube} to RDF data.
Yet, since existing approaches are adaptations of relational techniques, there is no understanding of their appropriateness to knowledge graphs.
We shed a light on the use of multiple alternative approaches over KGs by showcasing \oursystem, a system that compares various cost models for view materialization.
A cost model is the main building block for selecting the views to materialize, as it provides an estimate of the time for querying a database with and without the materialized views.

\mpara{Contributions.}
{\oursystem} proposes, evaluates, and compares a variety of existing cost models for view selection, adapted for the RDF setting. It allows users to run a set of queries on the materialized views and inspect the performance in executing the query workload.
The goal of this prototype is to identify strengths and limitations of multiple cost estimation techniques for view selection on RDF data.
In summary, {\oursystem}
(1) addresses the problem of providing fast query answering for analytical queries on KGs,
(2) provides a generic solution to be deployed on any RDF triple store with SPARQL query processing, and
(3) highlights possible limitations of six alternative approaches.
Given a KG, a \template over the KG, and a constraint on the number of views to materialize, \oursystem generates a set of views to answer aggregated queries over the provided \template.

%% file: sections/2.related.tex
\section{Related Works}\label{sec:related}

KGs gained traction in the last few years, due to the proliferation of Linked Open Data~\cite{Bonifati2019,wylot2018,seaborne2006sparql} and proprietary enterprise knowledge graphs~\cite{noy2019industry}.
Recently, companies and researchers require to perform complex analytics on the data in the form of aggregate queries.

In the following, we provide more details around existing methods for data cube analysis for the relational model and the existing implementations for the case of graph data.
We highlight how existing methods have tried to adapt techniques for relational data to the graph model.
In this demonstration, we present a system that can showcase the limitations of these adaptations.

\mpara{Data cube analysis.}
In relational data, \emph{data cubes}~\cite{olapcube} conveniently represent aggregates over multiple data dimensions.
That is, they model data as a set of observations, each carrying one or more measures, and a set of dimensions across which the measures of the observations can be aggregated (e.g., consider the population recorded for each city in each country, which can be aggregated across time, regions and continents, or language spoken in order to retrieve, for instance, the amount of population per country speaking each language).
Analyses in such data cubes are notoriously computationally expensive since they involve the processing of large portions of the dataset.
Therefore, a common approach is that of employing \emph{materialized views} so that queries can be executed over a smaller portion of pre-processed data, significantly reducing query time~\cite{olapcube,niemi2001constructing}.
For instance, one can pre-aggregate population across countries, languages, and years, so that a query asking for the total amount of people speaking German during 2020 can be computed by processing the pre-aggregated results instead of the whole data for each city.
Yet, given a data-cube with many different dimensions, there are multiple ways in which data could be aggregated (e.g., across cities and regions, or languages and years, and so on).
Materializing views for all these combinations is expensive both in terms of processing time as well as in terms of space occupation on disk.
Therefore, \emph{view selection} techniques have been proposed for the case of relational databases~\cite{olapcube,niemi2001constructing}.
These techniques estimate the benefit that materializing a specific view can provide.
Such benefit is estimated as a linear function of the size of the materialized view compared against the size of the data from which such a view should be derived.
For instance, a view aggregating daily records into yearly records provides an expected reduction factor of $\sim350$, and one would expect a proportional improvement in processing speed when using the view for querying, instead of the daily data.

For the case of RDF data, instead, the state of the art approaches simply set-out to adapt solutions from the relational model to the graph model.
Yet, the research on relational data cannot be directly applied on graphs, as the structure and the schema is not known a-priori in such datasets.

\mpara{OLAP approaches for RDF.}
The MARVEL system~\cite{olapcubeRDF}, belonging to this line of work, implements view materialization for optimizing query answering of OLAP SPARQL queries~\cite{etcheverry2012qb4olap}.
MARVEL employs a cost model, a view selection algorithm, and an algorithm for rewriting SPARQL queries using the available materialized views.
Although the approach is the first to tackle the challenges of answering analytical queries on KGs through view materialization, the input data should actually adopt a data cube model (in particular the QB4OLAP~\cite{etcheverry2012qb4olap}) and the cost model simply considers the number of edges (triples) in each view.

Other approaches have investigated the need for enabling complex aggregate queries in SPARQL~\cite{AnalyticsSPARQL,colazzo2014rdf}.
In particular, the Analytical schema model~\cite{colazzo2014rdf} enables different views on generic KGs.
Yet, this model does not tackle the problems of view materialization for RDF data, instead, they propose to map the data to a relational model and exploit traditional optimizations for relational queries.
Finally, a distinct approach for RDF analytics~\cite{AnalyticsSPARQL} converts a complex aggregate query to a set of smaller, approximate, queries.
Yet, this approach has the sole goal to diminish the load for the database answering the query, and not to speed up query processing.

Therefore, to date, no solution has explored in detail the case of view materialization for KGs as a graph-centric problem.
Instead, existing solutions, simply resort to map the data to a relational model.
{\oursystem} aims at systematically analyzing view materialization by shedding a light on existing methods to pave the road to a native graph-aware model for answering analytical queries on KGs.

%% file: sections/4.solution.tex
\section{The \oursystem System}\label{sec:solution}

\begin{figure}
    \centering
    \includegraphics[width=1.0\columnwidth]{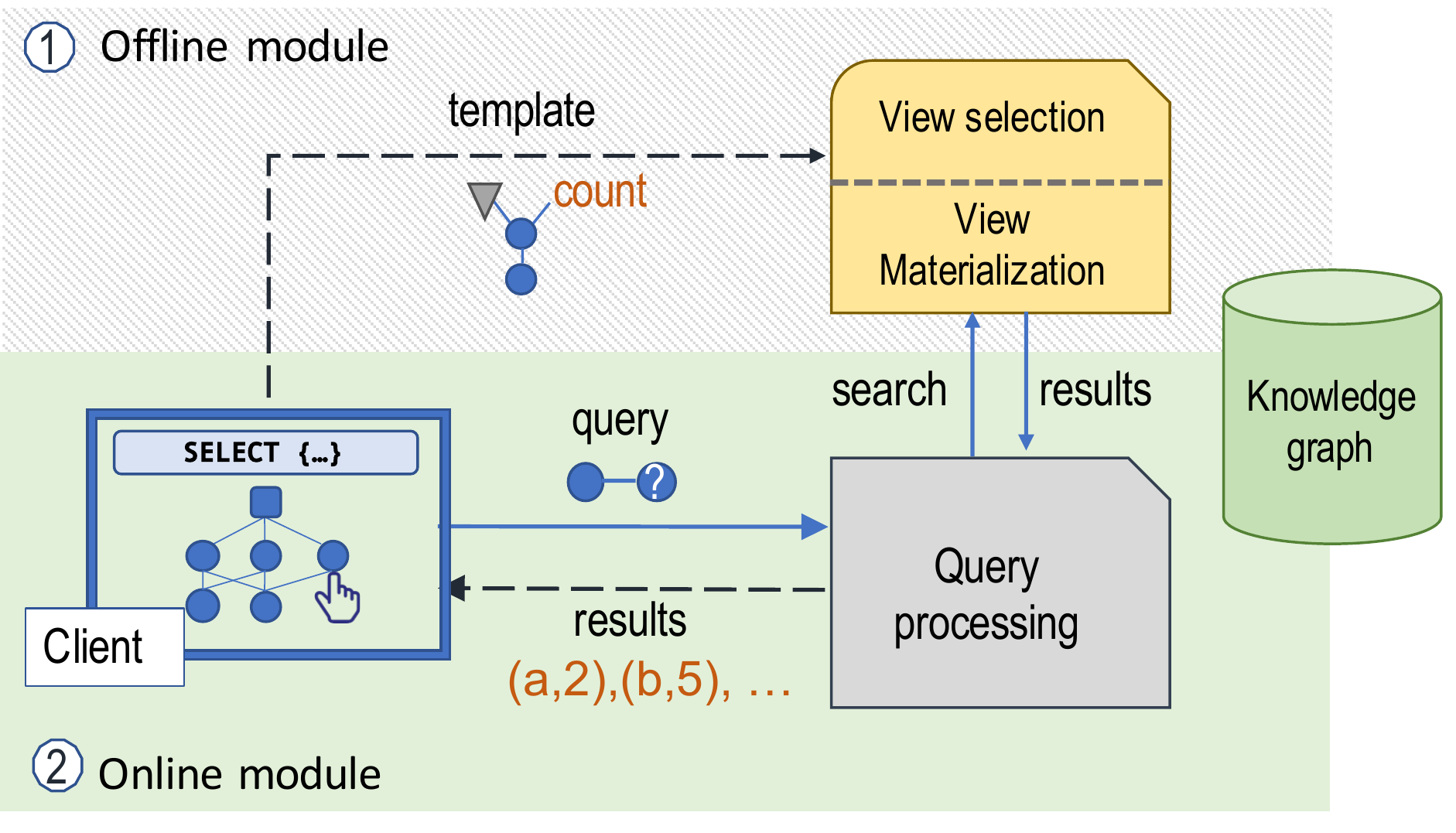}

    \caption{The \oursystem system.}
    \label{fig:our_system}

\end{figure}

\begin{figure*}[!ht]
    \centering
    \includegraphics[width=\textwidth]{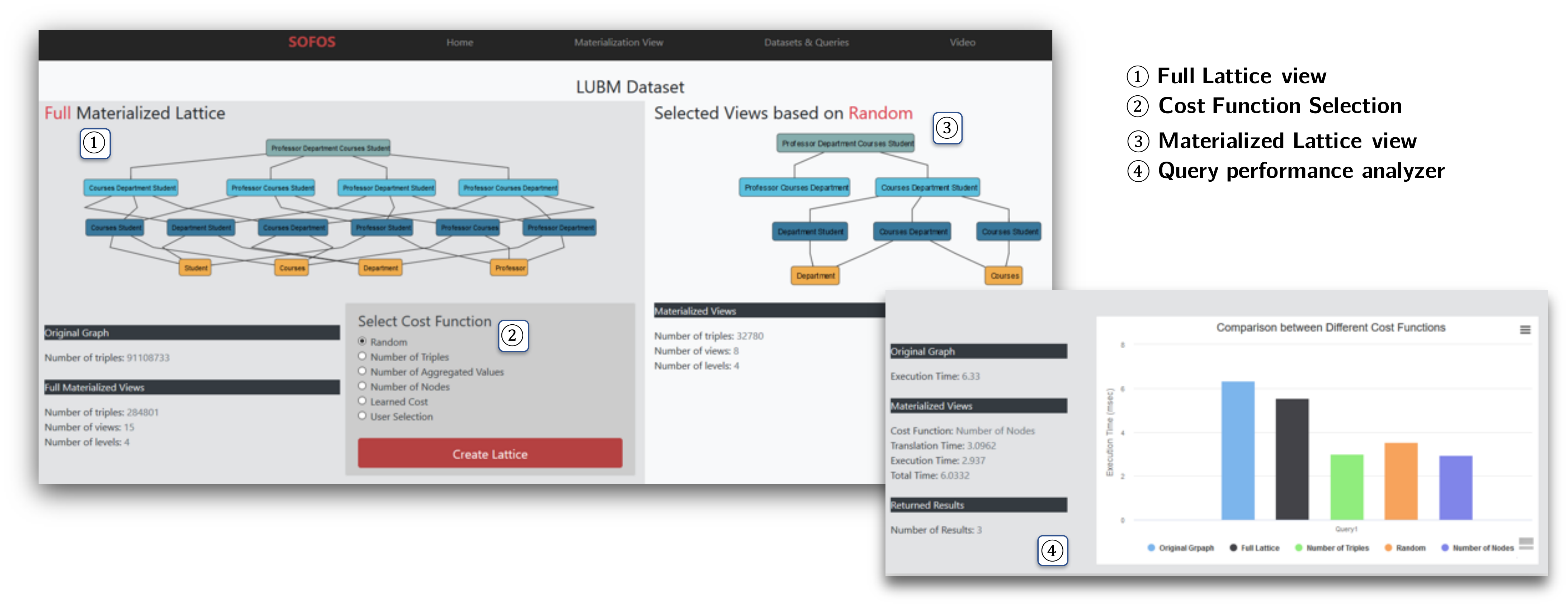}

    \caption{The GUI of \oursystem system.}
    \label{fig:gui}

\end{figure*}

The \oursystem system implements, adapts and compares several cost models for view selection on RDF KGs.
The system, given an initial analytical \emph{\template} of the graph to analyze, materializes a set of views based on a cost model, and then it measures the performance, in terms of storage cost and query response-time, of the selected views.
\oursystem comprises of two main modules:
{\large\ding{172}} an \textit{offline module} for \textbf{selective view materialization} (Section~\ref{ssec:offline}),
and
{\large\ding{173}} an \textit{online module} for \textbf{query execution and performance comparison} (Section~\ref{ssec:online}).
Figure~\ref{fig:our_system} shows its main components.

\mpara{Background \& problem:}
At its core, the \oursystem system takes a knowledge graph $\kg$ and an analytical \template $\T$, which describes the information that should be aggregated in different views, and materializes a set of $k$ views $\V_1, ..., \V_k$ based on $\T$.
Then, given any query $\Q$ targeting $\T$, the system either answers $\Q$ querying one of the $k$ materialized views, or accesses the graph $\kg$ if none of the views can be used to compute the required answer.

In \oursystem, a \emph{knowledge graph} $G$ is represented as a set of RDF triples $(s,p,o)\in(\iris\cup\blanks){\times}(\iris){\times}(\iris\cup\blanks\cup\literals)$, where $\iris$ is a set of entity identifiers, $\blanks$ is a set of ``blank'' nodes with no identifier, and $\literals$ is a set of literals. 
A \emph{query} $\Q$ on a RDF graph is a set of \emph{triple patterns}, that is, a set of triples in which some of the triple's components $s,p$, or $o$ are variables from a set $\vars$,
and is expressed in the SPARQL query language.
An \emph{answer} to a query $\Q$ is computed based on the  matchings in $\kg$ of the triple patterns in the query and the values corresponding to instances of the variables in the query.
We denote as $\res{\Q}{\kg}$ the set of query answers on the knowledge graph $\kg$.
Here, we focus on \emph{analytical queries} of the kind
\texttt{SELECT} $\varsq$~$agg(u)$ \texttt{WHERE} $P$ \texttt{GROUP BY} $\varsq$, in which $\varsq{\subseteq}\vars$ are grouping variables, i.e.,  a subset of the variables appearing in $P$, $u{\in}\varsq$ is the specific variable over which the aggregation is computed, and $agg$ is an aggregation expression in $\{ \textsc{SUM, AVG, COUNT, MAX, MIN} \}$.

The \oursystem system builds on analytical \emph{\templates} that determine the triples of the graph that are the target of some queries and hence provide the conditions to construct a set of views.
A \emph{\template} has the same form of an analytical query and is then identified by the triple $\T{=}{\langle}\varsq,P,agg(u){\rangle}$.
Finally, a \emph{view} from a \template $\T$ is a query $\V{=}{\langle}\varsq^{\prime},P^{\prime},agg(u){\rangle}$, where $P^{\prime}$ is derived from $P$, and $\varsq^{\prime}{\subseteq}\varsq$ aggregates over just a subset of variables in $\varsq$.
Therefore, the \template $\T$ induces a \emph{lattice} of views $\lattice{\T}$, in which different subsets of variables are used for aggregation and hence, results are represented at different levels of granularity.
Moreover, in {\oursystem}, a materialized view is also an RDF graph that contains an encoding of only the answers to the query used to generate it.
Analytical queries targeting a {\template} $\T$ also contain a subset of $\varsq$ and $P$ but can be further specialized by also introducing additional \texttt{FILTER} conditions.

Given a query $\Q$, \emph{view materialization} allows for answering the query by exploiting the contents of a precomputed view $\V_i$, avoiding in this way the need to query the underlying graph $\kg$.
Materializing the entire lattice would allow to always select the best view $V_i$ for any query.
Nonetheless, materializing the entire lattice is impractical from the memory consumption standpoint.
As such, \oursystem explores different strategies that have been proposed in the past to select a subset $\V_1, ..., \V_k$ of views from the lattice.
In the relational case, the system would always select the smallest possible view to answer $\Q$, since there is a linear correlation between number of tuples and running time~\cite{olapcube}.
This linear correlation does not trivially hold in the case of knowledge graphs, because a graph is not defined in terms of tuples.
As such, we need a cost function $C: \lattice{\T} \rightarrow \reals^+$  predicting the running time of any query Q if the view $V_i$ is materialized.

In practice, to select the best set of views, we adopt a greedy approach~\cite{olapcube}.
Given a set of selected views, the greedy approach exploits the estimated time from the cost function and compares the expected running time of a set of queries with and without including the candidate view $\V_i$ in the set of views.
While, in the relational case, the cost is derived directly from the number of tuples in the view,
\oursystem proposes a comparison among different cost functions to select $k$ views from a facet $\T$, and shows the advantages and shortcoming of each of them when tested against a specific set of queries.
We opt for a budget representing the number of views $k$ to allow for a more straightforward comparison on memory and time consumption.
However, note that this budget can be adapted to regulate the space consumption on the selected views as well, i.e., instead of selecting $k$ views, select up to $k$ views up to a certain memory budget.

\vspace{-10pt}
\subsection{Selective view materialization}\label{ssec:offline}

\oursystem performs two offline operations, (a) \textbf{view selection} that decides on the best views to materialize given the cost function $\C$, and (b) \textbf{view materialization} that augments the graph with extra information to store aggregation values.

\spara{View selection.} \oursystem supports six cost models: (1) a random baseline, (2) a direct adaptation of tuple counting for relational data, two RDF-based cost models, namely (3) the number of aggregated values and (4) the number of nodes, (5) a learned cost model, and (6) a user-defined one. 

\begin{itemize}[leftmargin=*]
    \item \textbf{Random:} This cost function is constant $\C(\V_i){=1}$, for each view $\V_i{\in}\lattice{\T}$, i.e., this will output a random $k$-size subset of $\lattice{\T}$.

    \item \textbf{Number of triples:} This cost function is analogous to the number of tuples in relational databases.
    On a knowledge graph, this cost corresponds to the number of RDF triples in the corresponding graph $G_{V_i}$, $\C(\V_i){=}|{G_{V_i}}|$.
    \item \textbf{Number of aggregated values:} This corresponds to the number of results of the query representing the view, i.e., $\C(\V_i){=} |\res{\V_i}{\kg}|$.
    \item \textbf{Number of nodes:} This cost corresponds to the number of node values in the view $\V_i$, i.e., $\C(\V_i){=}|\iris_i{\cup}\blanks_i{\cup}\literals_i|$.
    \item \textbf{Learned cost}: For comparison, we adapt a cost estimate from a learned deep regression model $f{:} \lattice{\T}{\rightarrow}\reals$~\cite{ortiz2019empirical}.  We encode a query into a vector representing the relationships, the attributes, and the type of aggregates in the query, along with statistics about the relationship frequency and the attribute frequency. In the offline training phase, the model takes the encoding of either a given workload or randomly generated queries and their running time. In the online phase, the model receives the encoding of a query (i.e., view) $V_i$ and outputs the estimated running time, such that $\C(\V_i) = f(\V_i)$.

    \item \textbf{User defined}: The user acts as a cost function, selecting $k$ views from the lattice.

\end{itemize}

\spara{View materialization.}
View materialization in {\oursystem} consists of generating a new graph for each view $\V_i$.
Each graph contains a set of extra blank nodes to which is attached the value of the aggregation of different bindings for the subset of the template variables in $\varsq$.
This materialization procedure is a generalization of the standard techniques adopted in MARVEL~\cite{olapcubeRDF}.
The result of view materialization is hence an \emph{expanded RDF graph} $\kg^+$.

\subsection{Query Performance Comparison}\label{ssec:online}

After materialization of a specific subset of views, the system runs a set of queries randomly generated from the {\template} $\T$ against the expanded graph $\kg^+$ and measures the performance of each query.

When answering a query, \oursystem identifies the best view to adopt and translates the input query $\Q$ into a query $\Q'$ in the expanded RDF graph $\kg^{+}$ targeting the data of the selected view.
In practice, the translation straightforwardly substitutes aggregate variables with the blank nodes representing the aggregation and reformulates triples patterns accordingly.

Therefore, \oursystem allows running any set of queries on different sets of materialized views for each cost function.
The user can then compare the relative performance of each view selection method and hence the appropriateness of different cost models.

%% file: sections/5.scenario.tex
\section{Demonstration Scenario}\label{sec:scenario}

The goal for the demonstration is to show, through experiments, the challenges involved in materialized view selection on knowledge graphs, exploring various alternative cost models.
A screenshot of our system is shown in Figure~\ref{fig:gui}.
The demonstration will start by guiding the participants through the different design choices in \oursystem.
We will then walk them through the following steps:

 \textbf{Configuration}: In this step, the three datasets used for our demonstration (i.e., the LUBM, the DBpedia, and the Semantic Web Dogfood datasets) will be presented along with the corresponding query \templates for these datasets.
 Each query \template will be accompanied by a high-level description and a corresponding SPARQL query template, enabling the active exploration of the data available each time.
 For each dataset we will propose a query workload composed of different parametrized queries for a given query template.

\textbf{Exploration of the Full Lattice:} By selecting a specific combination of dataset and \template, the full materialized lattice will be presented to the users, explaining why such a large structure is required, precomputing at the various levels, the aggregations that the query template might ask.
By selecting a node (view) in the lattice the user will be able to check the data that are stored for this specific node.

\textbf{Exploring Cost Models:} Using the full lattice as input, the various view selection algorithms (and the accompanied cost models) will be explained to the participants and demonstrated in practice.
In each case, the trade-off in query execution and storage amplification will be shown, enabling users to understand which cost model is better in each case.

\textbf{User Selected Views:} Besides exploiting an already existing view selection algorithm, the users will be able to select individual nodes from the lattice to be materialized and see the impact of their choices on the query execution time.
Each time the space amplification and the query execution time will be contrasted, enabling users to explore the sweet-spot where space amplification is minimized and query execution time is improved.

\textbf{``Hands-on'' Challenge:}
In this phase, conference participants would be challenged, so that given a specific query and budget, to optimally select the views to be materialized for optimizing query execution.
The participant that will make the best selection will receive a {\oursystem}-related small prize.